\documentclass[conference,letterpaper]{IEEEtran}
\IEEEoverridecommandlockouts

\usepackage{fancyhdr}
\usepackage[dvips]{graphicx}
\usepackage{indent}
\usepackage{amsmath,amssymb}
\usepackage{paralist}
\usepackage{eclbkbox}
\usepackage{subfigure}

\begin{document}
\title{A Clonal Selection Algorithm with Levenshtein Distance based Image Similarity in Multidimensional Subjective Tourist Information and Discovery of Cryptic Spots by Interactive GHSOM
\thanks{\copyright 2013 IEEE. Personal use of this material is permitted. Permission from IEEE must be obtained for all other uses, in any current or future media, including reprinting/republishing this material for advertising or promotional purposes, creating new collective works, for resale or redistribution to servers or lists, or reuse of any copyrighted component of this work in other works.}}

\author{\IEEEauthorblockN{Takumi Ichimura}
\IEEEauthorblockA{Faculty of Management and Information Systems,\\
Prefectural University of Hiroshima\\
1-1-71, Ujina-Higashi, Minami-ku,\\
Hiroshima, 734-8559, Japan\\
Email: ichimura@pu-hiroshima.ac.jp}
\and
\IEEEauthorblockN{Shin Kamada}
\IEEEauthorblockA{Graduate School of Comprehensive Scientific Research,\\
Prefectural University of Hiroshima\\
1-1-71, Ujina-Higashi, Minami-ku,\\
Hiroshima, 734-8559, Japan\\
Email: shinkamada46@gmail.com}
}

\maketitle
\thispagestyle{plain}

\fancypagestyle{plain}{
\fancyhf{}	
\fancyfoot[L]{}
\fancyfoot[C]{}
\fancyfoot[R]{}
\renewcommand{\headrulewidth}{0pt}
\renewcommand{\footrulewidth}{0pt}
}

\pagestyle{fancy}{
\fancyhf{}
\fancyfoot[R]{}}
\renewcommand{\headrulewidth}{0pt}
\renewcommand{\footrulewidth}{0pt}

\begin{abstract}
Mobile Phone based Participatory Sensing (MPPS) system involves a community of users sending personal information and participating in autonomous sensing through their mobile phones. Sensed data can also be obtained from external sensing devices that can communicate wirelessly to the phone. Our developed tourist subjective data collection system with Android smartphone can determine the filtering rules to provide the important information of sightseeing spot. The rules are automatically generated by Interactive Growing Hierarchical SOM. However, the filtering rules related to photograph were not generated, because the extraction of the specified characteristics from images cannot be realized. We propose the effective method of the Levenshtein distance to deduce the spatial proximity of image viewpoints and thus determine the specified pattern in which images should be processed. To verify the proposed method, some experiments to classify the subjective data with images are executed by Interactive GHSOM and Clonal Selection Algorithm with Immunological Memory Cells in this paper.
\end{abstract}


\IEEEpeerreviewmaketitle

\section{Introduction}
The current information technology can collect various data sets because the recent tremendous technical advances in processing power, storage capacity and network connected cloud computing. The sample record in such data set includes not only numerical values but also language, evaluation, and binary data such as pictures. The technical method to discover knowledge in such databases is known to be a field of data mining and developed in various research fields.

Mobile Phone based Participatory Sensing (MPPS) system involves a community of users sending personal information and participating in autonomous sensing through their mobile phones \cite{Lane2010}. Sensed data can be obtained from sensing devices present on mobiles such as audio, video, and motion sensors, the latter available in high-end mobile phones. Sensed data can also be obtained from external sensing devices that can communicate wirelessly to the phone. Participation of mobile phone users in sensorial data collection both from the individual and from the surrounding environment presents a wide range of opportunities for truly pervasive applications. The tourist subjective data collection system with Android smartphone has been developed \cite{Android_Market}. The application can collect subjective data such as pictures with GPS, geographic location name, the evaluation, and comments in real sightseeing spots where a tourist visits and more than 500 subjective data are stored in the database. Attractive knowledge discovery for sight seeing spots is required to promote the sightseeing industries. 

We have already proposed the classification method from the collected subjective data by the interactive GHSOM \cite{Ichimura12b,Ichimura12c} and the knowledge is extracted from the classification results of the interactive GHSOM by C4.5 \cite{Quinlan96}. However, the image data was not included in the classification tasks, because it is too large amount of information to realize the extraction of specified characteristics from images.

There is currently an abundance of vision algorithms which are capable of determining the relative positions of the viewpoints from which the images have been acquired. However, very few of these algorithms can cope with unordered image sets for which no a prior proximity ordering information is available. Image localization can be addressed in the framework of the fundamental structure and motion (SaM) estimation problem and benefits from the wide field of view offered by Smartphone camera. This is because a wide field of view facilitates capturing large portions of the environment with few images and without resorting to the use of movable gaze control mechanisms such as pan-tilt units. Furthermore, environment features remain visible in large subsets of images and critical surfaces are less likely to cover the whole visual field.

The definition of relative positions and orientations of the viewpoints corresponding to a set of unordered central images is an important procedure to be statistical analysis in image retrieval. The idea in the proposed approach employs the Levenshtein distance\cite{Levenshtein} to deduce the spatial proximity of image viewpoints and thus determine the specified pattern in which images should be processed. Horizontal matching method for localizing unordered panoramic images has been proposed\cite{Michel10}. In the method, all images have been acquired from a constant height above a planar ground and operates sequentially by the Levenshtein distance. Our proposed method can process not only in horizontal matching but also in vertical matching. In this paper, the photographs are divided into some categories according to the similarity by the clonal selection algorithm with immunological memory cells before the classification by GHSOM.

The area of artificial immune system (AIS) has been an ever-increasing interested in not only theoretical works but applications in pattern recognition, network security, and optimization \cite{Castro1}, \cite{Dasgupta}. AIS uses ideas gleaned from immunology in order to develop adaptive systems capable of performing a wide range of tasks in various research areas. Gao indicated the complementary roles of somatic hypermutation (HM) and receptor editing (RE) and presented a novel clonal selection algorithm called RECSA model by incorporating the Receptor Editing method \cite{Gao}. The immunological memory which leads to a perception that an individual is immune to a particular agent is realized by the clustering of the generated antibodies\cite{Ichimura12a}.

The remainder of this paper is organized as follows. In Section \ref{sec:CSAIM}, the clonal selection theory with memory cells will be explained briefly. The idea about the antibody structure of images by Levenshtein Distance and experimental results are discussed in Section \ref{sec:Similarity_Levenshtein_Distance}. Section \ref{sec:interactiveGHSOM} describes the algorithm of interactive GHSOM and its interface tool. Section \ref{sec:experiments} explains the tourist subjective data and the experimental results. In Section \ref{sec:ConclusiveDiscussion}, we give some discussions to conclude this paper.

\section{Clonal Selection Algorithm with Immunological Memory}
\label{sec:CSAIM}
Clonal Selection Algorithm with Immunological Memory(CSAIM) model has been proposed to introduce an idea of immunological memory into the RECSA model. This section describes the structure of antibody in RECSA model to the medical diagnosis briefly. The further details about the CSAIM algorithm was described in \cite{Ichimura12a}.

\subsection{Antibody for Classification Problem}
This subsection describes the antibody for classification problem about the structure, the method of somatic hypermutation and receptor editing, and affinity.
\subsubsection{Structure of Antibody for Classification Problem}
\label{sec:structure}
\begin{figure}[tbp]
\begin{center}
\includegraphics[scale=0.6]{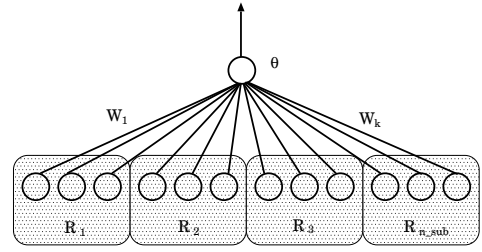}
\vspace{-3mm}
\caption{The antibody structure}
\label{fig:immune_individual}
\end{center}
\end{figure}

Fig.\ref{fig:immune_individual} shows the structure of antibody in the classification problem\cite{Ichimura12a}. $w_{k}$, $\theta$ is the weight of antibody and threshold, respectively. $R_{1}, \cdots, R_{n_{sub}}$ indicate the sub-region in the problem, because some classification problem can be divided into $n_{sub}$ sub tasks. That is, a region is expert for the specified task in classification. 

\begin{figure}[tbp]
\begin{center}
\includegraphics[scale=0.4]{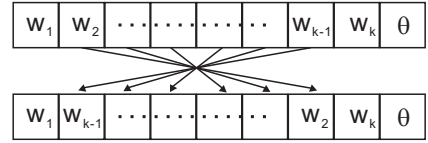}
\vspace{-3mm}
\caption{RE for $w_2,w_{k-1}$}
\label{fig:network_gene}
\end{center}
\end{figure}

\begin{figure}[tbp]
\begin{center}
\includegraphics[scale=0.5]{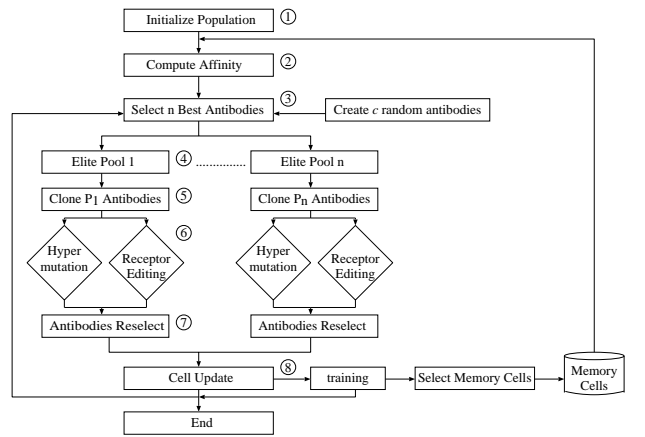}
\caption{A flow of CSAIM model}
\vspace{-3mm}
\label{fig:CSAIMmodel}
\end{center}
\end{figure}

\subsubsection{Somatic Hypermutation and Receptor Editing}
\label{sec:HMRE}
HM updates the randomly selected $w_i$ and $\theta$ for a paratope $P=(w_1,...,w_k,\theta)$ as follows.
\[ w_i=w_i+\Delta w, \theta=\theta+\Delta \theta, \]
where $\Delta w$, $\Delta \theta$ are $-\gamma_{w}<\Delta w<\gamma_{w}$, $-1<\Delta \theta <\gamma_{\theta}$, respectively. $\gamma_{w}$ and $\gamma_{theta}$ are a small number.

RE makes a crossover of 2 set of $w_i$ for a paratope as shown in Fig.\ref{fig:network_gene}.

\subsubsection{Affinity}
\label{sec:Affinity}
The system calculates the degree of affinities between antibody and antigen by using Eq.(\ref{eq:affinity}) and Eq.(\ref{eq:affinity2}).
\begin{equation}
f(x^p)=\left \{
\begin{array}{l l}
1 & if\ | \sum_{i=1}^{k} w_ix_{i}^{p} -\theta | \geq E_{sim}\\
0 &  otherwise
\end{array}
\right.
\label{eq:affinity}
\end{equation}

\begin{equation}
g(x^p)=\left \{
\begin{array}{l l}
1 & if \ \ f(x^p)=x_{Target}^p\\
0 &  otherwise
\end{array}
\right.
\label{eq:affinity2}
\end{equation}

Eq.(\ref{eq:affinity3}) calculates the degree of affinity.

\begin{equation}
\label{eq:affinity3}
Affinity=\sum_{p=1}^{tr\_num} g(x^p), 
\end{equation}
where $x^p$ means the $p$th sample in $tr\_num$ training cases and $x_{Target}^p$ is 1 if an example is a Target, otherwise 0.

\subsection{RECSA Model\cite{Gao}}
The shape-space model aims at quantitatively describing the interactions among {\it Ag}s and {\it Ab}s ({\it Ag}-{\it Ab}) \cite{Perelson}. The set of features that characterize a molecule is called its generalized shape. The {\it Ag}-{\it Ab} codification determines their spatial representation and a distance measure is used to calculate the degree of interaction between these molecules. 

The Gao's model \cite{Gao} can be described as follows.
\begin{description}
\setlength{\itemindent}{0.0cm}
\setlength{\labelsep}{1pt}
\begin{indentation}{-1.0cm}{0.0cm}

\item[1)] Create an initial pool of $m$ antibodies as candidate solutions$({\it Ab_{1}}, {\it Ab_{2}}, \cdots, {\it Ab_{m}})$.

\item[2)] Compute the affinity of all antibodies: $(D({\it Ab_{1}}), D({\it Ab_{2}}), \cdots, D({\it Ab_{m}}))$. $D()$ means the function to compute the affinity.

\item[3)] Select $n$ best individuals based on their affinities from the $m$ original antibodies. These antibodies will be referred to as the elites.

\item[4)] Sort the $n$ selected elites in $n$ separate and distinct pools in ascending order. They will be referred to as the elite pools.

\item[5)] Clone the elites in the pool with a rate proportional to its fitness. The amount of clone generated for these antibodies is given by Eq.(\ref{eq2-1}).
\vspace{-3mm}
\begin{equation}
P_{i} = round(\frac{n-i}{n}\times Q ),
\label{eq2-1}
\end{equation}
where $i$ is the ordinal number of the elite pools, $Q$ is a multiplying factor for determining the scope of the clone and $round()$ is the operator that rounds towards the closest integer. Then, we can obtain $\sum P_{i}$ antibodies as $(({\it Ab_{1,1}}, {\it Ab_{1,2}}, \cdots,$ ${\it Ab_{1,p_{1}}}), \cdots, ({\it Ab_{n,1}}, {\it Ab_{n,2}}, \cdots, {\it Ab_{n,p_{n}}}))$.

\item[6)] Subject the clones in each pool through either hypermutation or receptor editing process. The mutation rates, $P_{hm}$ for hypermutation and $P_{re}$ for receptor editing given by Eq.(\ref{eq2-2}) and Eq.(\ref{eq2-3}), are inversely proportional to the fitness of the parent antibody,
\begin{equation}
P_{hm} = a/D()
\label{eq2-2}
\end{equation}
\begin{equation}
P_{re} = (D()-a)/D(),
\label{eq2-3}
\end{equation}
where $D()$ is the affinity of the current parent antibody and $a$ is an appropriate numerous value.

\item[7)] Determine the fittest individual $B_{i}$ in each elite pool from amongst its mutated clones. The $B_{i}$ is satisfied with the following equation.
\begin{eqnarray}
\nonumber D(B_{i}) &=& max(D({\it Ab_{i,1}}), \cdots, D({\it Ab_{i,p_{i}}})),\\
i&=&1,2,\cdots,n
\label{eq3-4}
\end{eqnarray}

\item[8)] Update the parent antibodies in each elite pool with the fittest individual of the clones and the probability $P({\it Ab_{i}}, \rightarrow B_{i})$ is according to the roles: if $D({\it Ab_{i}}) < D(B_{i})$ then $P=1$, if $D({\it Ab_{1}}) \ge D(B_{1})$ then $P=0$, if $D({\it Ab_{i}}) \ge D(B_{i}), i \ne 1$ then $P=exp(\frac{D(B_{i})-D({\it Ab_{i}})}{\alpha})$.

\item[9)] Replace the worst $c$($=\beta \times n$, $\beta$ is the parameter.) elite pools with new random antibodies once every $t$ generations to introduce diversity and prevent the search from being trapped in local optima. 

\item[10)] Determine if the maximum number of generation $G_{max}$ to evolve is reached. If it is satisfied with this condition, it terminates and returns the best antibody. Otherwise, go to Step 4).
\end{indentation}
\end{description}

\subsection{Immunological Memory Cell}
\label{sec:IMC}

\begin{figure}[tbp]
\begin{center}
\includegraphics[scale=0.5]{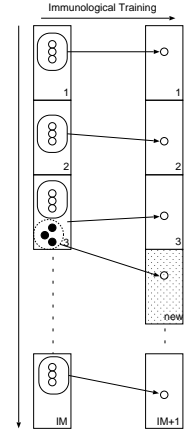}
\caption{A clustering method of memory cells}
\vspace{-5mm}
\label{fig:CSAIMmodel_clustering}
\end{center}
\end{figure}

Clustering Memory Cells are required to classify the antibodies responding the specified samples. This paper realizes the clustering by allocating the generated antibodies by RECSA model into some categories. The initial number of categories is predefined and a new category is created according to training situation. Fig. \ref{fig:CSAIMmodel_clustering} shows the clustering method of memory cells. Similar antibodies crowd around an appropriate point in each category, and then only central antibody of the crowd can become a memory cell. However, we may meet that memory cells can not recognize some of samples in the data set. In such a case, some new generated antibodies by RECSA model tries to respond to the mis-classification of the samples, if the similar antibodies make a crowd.

To find the crowd of similar cases, the system checks whether the Euclidean distance between normalized training sample and its corresponding antibody is smaller than the predetermined parameter $\mu_{\theta}$.

The similarity is measured by the following. Let $\vec{d}=(d_1,\cdots,d_i,\cdots,d_k)$ be the elements of input signal and $\vec{h}=(h_1,\cdots,h_i,\cdots,h_k)$ be the element of antibody.

In order to calculate the distance between the sample and the antibody, the range of sample is changed to that of antibody as follows.
\vspace{-3mm}
\begin{equation}
\nonumber d_{i}^{'}=d_{i} \times \frac{h_{j}}{d_{j}} (d_{i} \neq  0 \wedge h_{i} \neq 0),
\end{equation}
where $d_{j}$ is the min value of element in the input sample.

Then, if the Euclidean distance between $\vec{d^{'}}$ and $\vec{h}$ is smaller than $\mu_{\theta}$, the antibody can respond the sample. In this paper, $\mu_{\theta}$ is the summation of 12 input elements.

\section{Similarity of Images by Levenshtein Distance}
\label{sec:Similarity_Levenshtein_Distance}
\subsection{Levenshtein Distance}
\label{sec:Levenshutein_Distance}
The Levenshtein Distance (LD) is a string metric for measuring the difference between two sequences, the source ($s$) and target ($t$). The LD between $s$ and $t$ corresponds to the minimum number of edits needed to transform one string into the other. The allowable operation to edit are defined as letter deletion, insertion, and substitution of a single character. The LD can be computed in $O(|s||t|)$ time by a dynamic programming technique, known as the Levenshtein algorithm. That is, LD between $s$ and $t$, $lev_{s,t}(|s||t|)$, is given as follows.
\vspace{-3mm}
\begin{eqnarray}
lev_{s,t}(i,j)=\left\{
\begin{array}{l}
\max(i,j)\hfill, {\rm if} \min(i,j)=0\\
\min \left \{ \begin{array}{l}lev_{s,t}(i-1,j)+1,\\lev_{s,t}(i,j-1)+1,\\lev_{s,t}(i-1,j-1)+[s_{i}\neq t_{j}])\\ \end{array} \right.
,\\ \hfill, otherwise\\
\end{array}
\right.
\label{eq:Levenshtein}
\end{eqnarray}

As a byproduct, this algorithm returns the pairs of letters that have been matched while computing the LD. If a letter at position $i$ in $s$ matches the letter at position $j$ in $t$, then letters in $s$ at positions $k > i$ can only match letters in $t$ that are at positions $l > j$. The LD has been employed in various domains in need of pattern matching, such as spell checking, pattern recognition, speech recognition, information theory, cryptology, bio informatics, and so on. With respect to the field of computer vision, the use of the LD has been rather limited and has concerned the comparison of graph structures under edit operations\cite{Ng2002}.

\subsection{Data Sequence in Photo Images}
\label{sec:datasequence}
The matching method using Levenshtein distance described in \cite{Michel10} is the recovering the position and orientation parameters corresponding to the viewpoints of a set of panoramic images. However, the method is effective only for the horizontal diminishing scale, that is, only for converting the panoramic image into the original image. There is no advantage to the scale for the usual photo image and the similarity between photo images.

\begin{figure}[tbp]
\begin{center}
\includegraphics[scale=0.35]{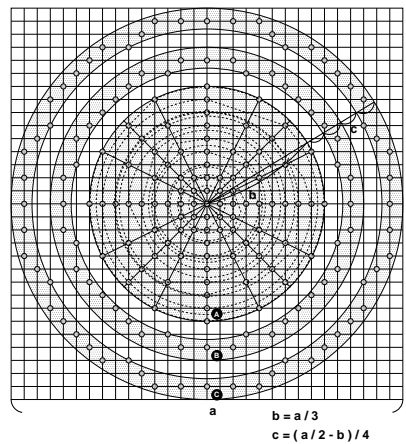}
\caption{A sequence of pixel in a photo image}
\vspace{-5mm}
\label{fig:LD_sequence}
\end{center}
\end{figure}

In this paper, the proposed method uses a newly-devised sequence of pixel in the usual photograph image as shown in Fig.\ref{fig:LD_sequence}. The grid in Fig.\ref{fig:LD_sequence} means the pixel in the focused image, which is the main person or the main scene in a picture. The symbol \textcircled{\scriptsize A}, \textcircled{\scriptsize B}, and \textcircled{\scriptsize C} in Fig.\ref{fig:LD_sequence} is 3 kinds of concentric rings. The circular segment of curves constituting tubes, \textcircled{\scriptsize B} and \textcircled{\scriptsize C} are located in $1/3$ and $2/3$ size of circle, respectively. The diameter of circle \textcircled{\scriptsize A} is $1/3$. The sequence of pixel in each region is  arranged in a spiral pattern spreading outward from a central source.

\begin{figure}[tbp]
\begin{center}
\includegraphics[scale=0.3]{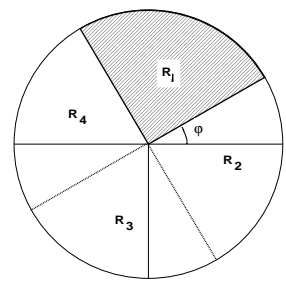}
\caption{4 sub-regions on roulette wheel}
\vspace{-5mm}
\label{fig:LD_region}
\end{center}
\end{figure}

Moreover, the circle is divided into 4 sub-regions on the roulette wheel and one of them is selected to measure the similarity of sequences between the photo image and the representative image such as a landmark in the famous spot. The fiabellate shape is randomly chosen to measure similarity in many times. $\psi$ in Fig.\ref{fig:LD_region} is an arbitrary gap from a rotation axis. The similarity is classified by CSAIM in the section \ref{sec:IMC}.

\section{Interactive GHSOM\cite{Ichimura12b, Ichimura12c}}
\label{sec:interactiveGHSOM}

\subsection{Growing Hierarchical SOM}
\label{sec:GHSOM}
\begin{figure}[tb]
\begin{center}
\includegraphics[scale=0.4]{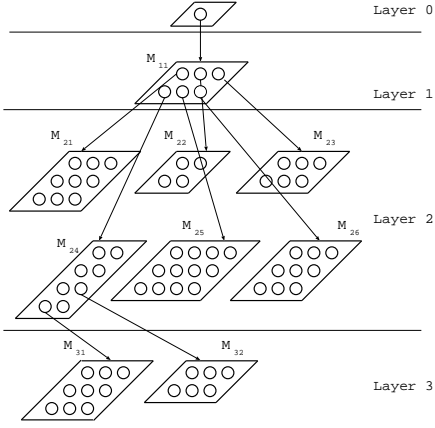}
\caption{A Hierarchy Structure in GHSOM}
\vspace{-3mm}
\label{fig:overviewGHSOM}
\end{center}
\end{figure}

A basic algorithm of GHSOM was described in \cite{Rauber02}. The algorithm has been chosen for its capability to develop a hierarchical structure of clustering and for the intuitive outputs which help the interpretation of the clusters. These capabilities allow different classification results from rough sketch to very detailed grain of knowledge. This technique is a development of SOM, a popular unsupervised neural network model for the analysis of high dimensional input data \cite{Kohonen95}. Fig.\ref{fig:overviewGHSOM} shows the overview of hierarchy structure in GHSOM.

\subsection{Control of growing hierarchies}
The process of unit insertion and layer stratification in GHSOM works according to the value of 2 kinds of parameters. Because the threshold of their parameters gives a criterion to hierarchies in GHSOM, GHSOM cannot change its structure to data samples adaptively while training maps. Therefore, only a few samples are occurred in a terminal map of hierarchies. In such a case, GHSOM has a complex tree structure and many nodes(maps). As for these classification results, the acquired knowledge from the structure is lesser in scope or effect in the data mining. When we grasp the rough answer from the specification in the data set, the optimal set of parameters must be given to a traditional GHSOM. It is very difficult to find the optimal values through empirical studies.

We proposed the reconstruction method of hierarchy of GHSOM even if the deeper GHSOM is performed\cite{Ichimura12b, Ichimura12c}. A stopping criterion for stratification is defined. Moreover, if the quantization error is large and the condition of hierarchies is not satisfied, the requirements for redistribution of error are defined.

\begin{description}
\item[Case1)]If Eq.(\ref{eq:ichimuraGHSOM-1}) and the classification capability for hierarchies are satisfied, stop the process of hierarchies and insert new units in the map again.
\vspace{-3mm}
\begin{equation}
n_{k}\leq \alpha n_{I},
\label{eq:ichimuraGHSOM-1}
\end{equation}
where, $n_{k},\, n_{I}$ mean the number of input samples for the winner unit $k$ and of the all input samples ${\it I}$, respectively. The $\alpha$ is a constant.
\item[Case2)]If the quantization error is not larger and the addition of layer is not executed, we may meet the situation that the quantization error of a unit is larger than the quantization error in an overall map. If Eq.(\ref{eq:ichimuraGHSOM-2}) is satisfied, then a new unit is inserted.
\vspace{-3mm}
\begin{equation}
qe_{k}\geq \beta \tau_{1} \sum qe_{y},\, y \in {\bf S}_{k},
\label{eq:ichimuraGHSOM-2}
\end{equation}
where ${\bf S}_{k}$ is the set of winner units $k$. $\beta\tau_{1}$ is a constant for the quantization error.
\end{description}

\subsection{An interface of interactive GHSOM}
\begin{figure}[tbp]
\begin{center}
\includegraphics[scale=0.4]{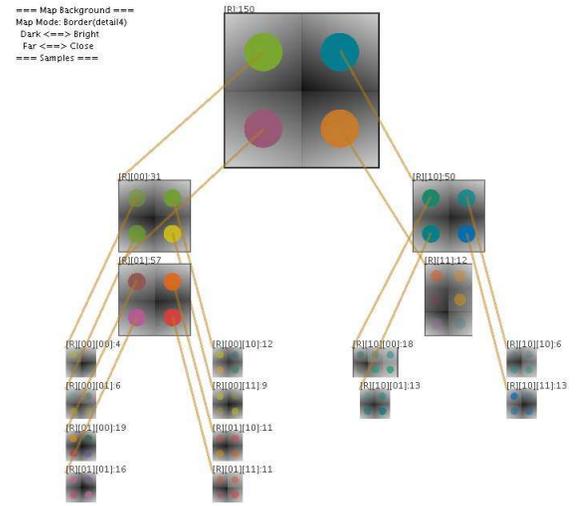}
\caption{Simulation Result for Iris data set}
\vspace{-3mm}
\label{fig:GHSOM_iris}
\end{center}
\end{figure}

We developed the Android smartphone based interface of interactive GHSOM to acquire the knowledge intuitively. This tool was developed by Java language. Fig. \ref{fig:GHSOM_iris} shows the clustering results of Iris data set \cite{UCI_IRIS} by GHSOM. 
The notation $[R][01][10]:11$ as shown in Fig. \ref{fig:GHSOM_iris} represents the location of unit in the connection from the top level $[R]$. $[R]$ means a root node. The numerical value('11') shows the number of samples divided into the leaf map after the sequence of classification $[R][01][10]$. The numerical values in the brackets mean the position of units in the corresponded map. The first letter (e.g. `0') and the second letter (e.g. `1') are the position in the column and the row in the map(e.g. `01'), respectively.

 The similar color of units represents an intuitive understanding of similar pattern of samples. If the number of units in a map are increased, only a few samples could be classified into a new generated unit. Once the unit connected to the map is selected, the method re-calculates to find an optimal set of weights in the local tree structure search and then a better structure is depicted.

When the corresponding unit is touched, the system calculates the 4 allocated samples. The system continues to classify again till the user determines the GHSOM structure. We call the method the interactive GHSOM in the interactive process. The calculation result by the interactive GHSOM as shown in Fig. \ref{fig:GHSOM_iris} is obtained and the effectiveness of the interactive GHSOM is shown as results of empirical studies.

The classification by GHSOM shows the tree structure of clusters and the connection among them. The detailed knowledge cannot be represented in the form of If-Then rules. Despite low resolution in knowledge representation, we grasp the rough sketch of knowledge structure because GHSOM shows the samples divided into each unit on the map as shown in Fig. \ref{fig:GHSOM_iris}. Moreover, there is only a few samples in each unit. Therefore, knowledge discovery is executed by grasping the structure and a grain of knowledge.

\section{Experimental Result}
\label{sec:experiments}
Participation of mobile phone users in sensorial data collection both from the individual and from the surrounding environment presents a wide range of opportunities for truly pervasive applications \cite{Lane2010}. Our developed Android smartphone application \cite{Android_Market} can collect the tourist subjective data in the research field of MPPS. The collected subjective data consist of jpeg files with GPS, geographic location name, the evaluation of $\{0, 1, 2, 3, 4\}$ and comments written in natural language at sightseeing spots to which a user really visits. The `comment' must be converted the number of words extracted from html files in the Tourist websites to a numerical value. The term frequency in the subjective comments is calculated by TF-IDF (term frequency inverse document frequency) method \cite{TF_IDF}. More than 500 subjective data are stored in the database through MPPS.

\begin{figure}[tbp]
\begin{center}
\subfigure[Without photos]{
\includegraphics[scale=0.38]{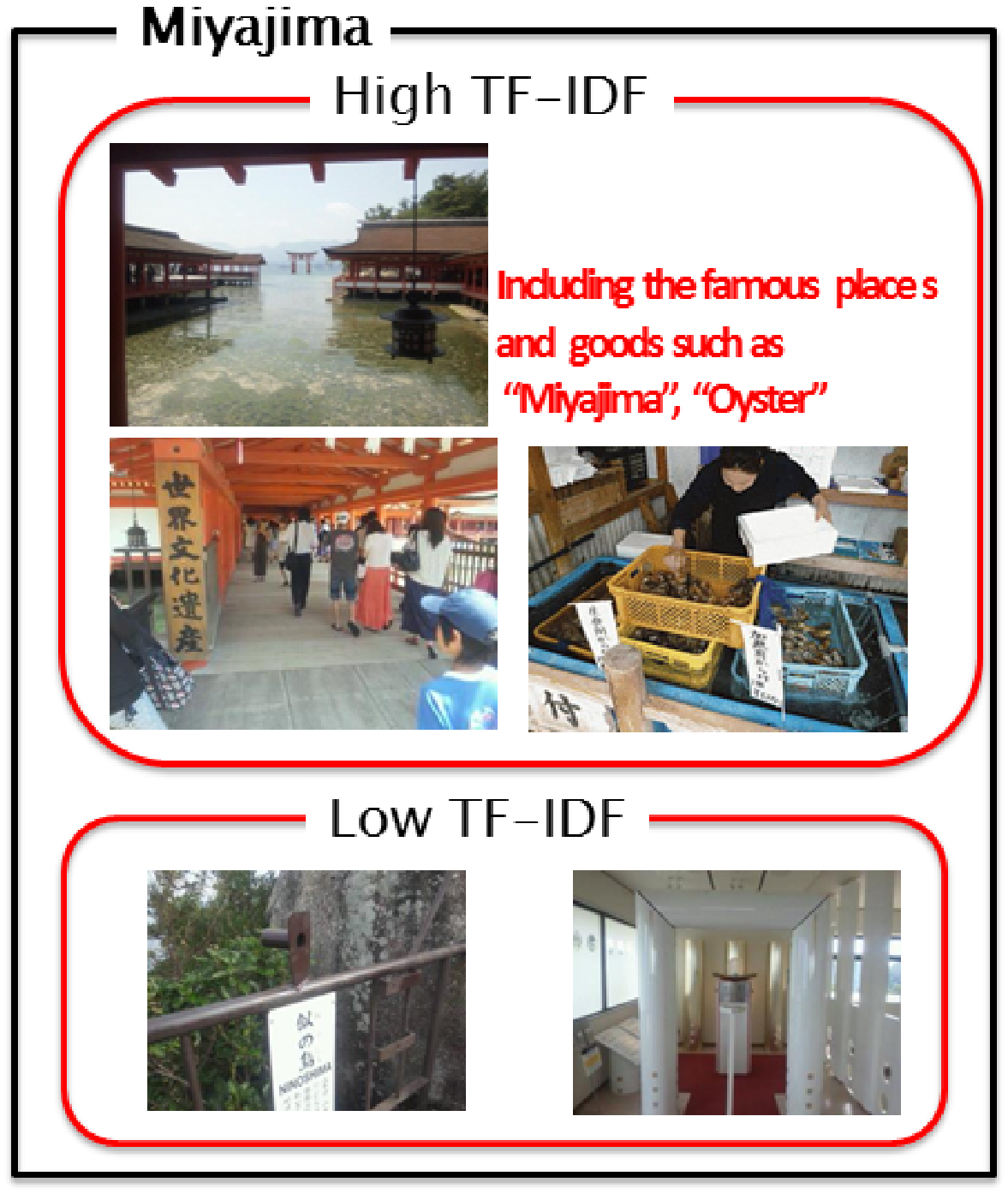}
\label{fig:Miyajima_wo}
}
\subfigure[With Levenshtein distance based similarity]{
\includegraphics[scale=0.35]{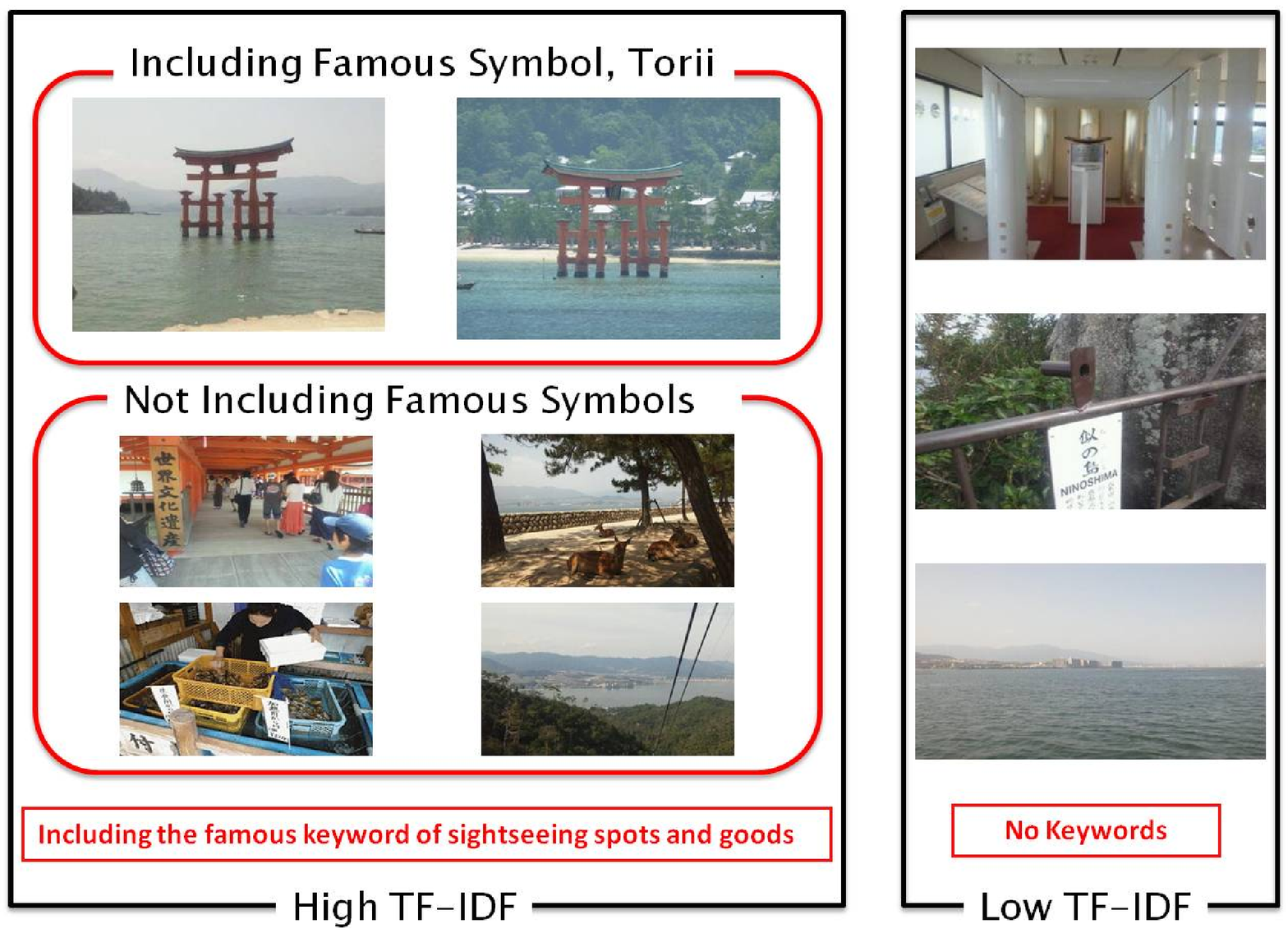}
\label{fig:Miyajima_w}
}

\vspace{-3mm}
\caption{Classification Result around Miyajima}
\label{fig:Miyajima}
\end{center}
\end{figure}

Fig.\ref{fig:Miyajima} shows the overview of classification result of subjective data by GHSOM. Fig.\ref{fig:Miyajima_wo} shows clusters by using only GPS, comments, and evaluation. The data without photos was classified by GHSOM, because the image data has a large amount of information and it is difficult for GHSOM to classify them as it is. On the contrary, Fig.\ref{fig:Miyajima_w} is the classification result with the image which was processed by using the Levenshtein distance based similarity of images. In the paper, we prepare the images for 6 representative spots. Especially, the famous symbol `Torii', a gateway at the entrance to a shrin, was drawn. The clusters in Fig.\ref{fig:Miyajima_w} are divided into 3 groups; `high similarity image with high TF-IDF and high evaluation', `low similarity image with high TF-IDF and high evaluation', and `low similarity image with low TF-IDF and low evaluation'. The data with `high similarity image' represents the famous sightseeing spot. The data with `high TF-IDF and high evaluation' and without `high similarity image' means that the spot is including the cryptic tourist information. Therefore, the subjective data has scarcity value and will be a sightseeing spot potentially. Moreover, the classification of image data could distinguish the unknown spot in the field of famous landmark. In other words, the cryptic spots will be discovered in near landmark.

\section{Conclusive Discussion}
\label{sec:ConclusiveDiscussion}
This paper presents an efficient and robust method for extracting the specified the images and arrange them in ascending order of the similarity, which represents the degree including the image of landmark. Matching a limited amount of image data has been shown to suffice for registering the images. CSAIM method can classify the specified pattern extracted from the images and then the spices in the pattern are divided into some groups. The tourist subjective data with the specified image which are collected through MPPS is classified by Interactive GHSOM. As a result, the discovery of cryptic sightseeing spots is executed. The similarity to the unfamiliar landmark in sightseeing spots does not measured. In order to improve such a problem, more subjective data is required to classify them and the social action among participants will be investigated in future.

\vspace{-3mm}
\section*{Acknowledgment}
This work was supported by JSPS KAKENHI Grant Number 25330366.

\end{document}